# Design and validation of a fuzzy logic controller for multi-section continuum robots

*Jing Liu, Tianyi Zeng, Abdelkhalick Mohammad, Xin Dong*, Dragos Axinte*

***Abstract*— The rise of multi-section continuum robots (CRs) has captivated researchers and practitioners across diverse industries and medical fields. Accurate modeling of these dexterous manipulators continues to be a significant challenge. This complexity stems primarily from many nonlinearities that plague their behavior, including hysteresis and cable elongation. Researchers have devised a spectrum of model-based and learning-based strategies to navigate this intricate landscape, aiming to conquer the modeling problem and elevate control performance. Despite the advancements in these approaches, they encounter challenges stemming from their complex design and intricate learning processes, impairing versatility and hindering robust closed-loop control. This paper introduces a simple-structured, model-less fuzzy logic controller for the closed-loop control of continuum robots. Unlike traditional methods relying on complex models and numerous sensors, this controller boasts a built-in shape reconstruction algorithm. This algorithm allows it to achieve robust control using only the feedback of end position and orientation, significantly reducing sensor dependence. It efficiently adapts to various nonlinearities like hysteresis, cable elongation, and unexpected external disturbances. The experimental results conclusively demonstrate the accuracy and robustness of the proposed fuzzy controller. On a three-section, six-degree-of-freedom continuum robot, it achieved a miniscule trajectory tracking Root Mean Square Error (RMSE) from 0.28 to 0.54 mm, representing just 0.17 to 0.32% of the robot's length. Additionally, the controller demonstrates robustness by successfully handling an unexpected external disturbance of 100g during the trajectory tracking.**

***Index Terms*—Closed-loop control, Continuum robot, Fuzzy logic controller, Robustness.**

## I. INTRODUCTION

CONTINUUM robots (CRs) have emerged as a rapidly growing field of interest due to their unique dexterity and maneuverability, making them promising tools for diverse applications such as industrial inspection[1], maintenance, deep sea exploration[2], and medical surgery[3][4]. For instance, in aerospace engine maintenance, CRs can thread end-effector tools through narrow inspection holes, reaching engine internals without disassembly [5], which is less time-consuming. Thanks to the unique separation of CRs' highly dexterous manipulator from the remotely located actuating system, they can be used as disposable tools in hazardous environments, minimizing contamination risks and costs. They are also ideal for minimally invasive surgery, and they can enhance patient safety and simplify interventions.

Continuum robots fall into three main categories: cable-driven CRs, concentric tube CRs, and soft CRs. Cable-driven continuum robots typically comprise one or more sections containing multiple vertebrae actuated by various driving cables [6][7]. Cable-driven CRs boast the highest degrees of freedom (DoFs) and the strongest load-carrying capacity among the three types. They typically feature a hollow interior for power and control cables routing to the end effector. In contrast, concentric tube CRs, consisting of a series of pre-bent Nitinol tubes, represent the thinnest category of CRs[8]. It is long and slender and has a smooth outer surface that allows easy access to tiny holes, which has been demonstrated as a promising solution for minimally invasive surgery. Unlike their counterparts, soft CRs embrace flexible materials like silicone, plastic film, and natural rubber. This versatility in material leads to diverse designs and, consequently, a wide range of actuation methods tailored to specific needs. While some soft CRs employ cable-driven mechanisms [9], others may utilize methods like pneumatic [10][11] and hydraulic [12].

Recent research has delved into both open-loop and closed-loop control strategies for CRs. Open-loop control hinges on precise kinematic modeling. The forward kinematics (FK) of a CR maps the configuration space to the task space[13]. Piecewise constant curvature (PCC) is a classical model of the CR forward kinematics [13][14]. This approach models a continuum robot as tangent continuous arcs, which is simple and computationally inexpensive. However, this approach neglects critical nonlinearities, including tendon friction and the backbone's torsional and elastic deformations. This results in a mismatch between the model and the actual shape of the robot [15]. The Cosserat rod model models the CRs as a deformable curve, which provides a geometrically accurate model but requires extensive calculations [16]. Inverse kinematics (IK) poses a more significant challenge than FK, necessitating more intricate algorithms. These algorithms fall into two primary categories: model-based and learning-based, differentiated by their reliance on a prior model of the continuum robot. Model-based methods typically leverage a

This work was supported by the University of Nottingham and Innovate U.K. through REINSTATE: Repair, Enhanced Inspection, and Novel Sensing Techniques for increased Availability and reduced Through life Expense under Project Reference under Grant 51689, China Scholarship Council (CSC). (Corresponding author: Xin Dong.)

Jing Liu, Tianyi Zeng, Abdelkhalick Mohammad, Xin Dong and Dragos Axinte are with the Department of Mechanical, Materials and Manufacturing Engineering, University of Nottingham, Nottingham NG7 2RD, U.K. (e-mail: eexjl54@nottingham.ac.uk; tianyi.zeng@nottingham.ac.uk; abd.mohammad1@nottingham.ac.uk; xin.dong@nottingham.ac.uk; dragos.axinte@nottingham.ac.uk).

forward kinematics model to compute iteratively numerical solutions to the IK problem. For example, a model-based method [15] was proposed to solve the inverse kinematics of a Magnetic Resonance Imaging(MRI)-guided magnetically actuated single-section catheter system, utilizing the inverse Jacobian matrix. This approach achieves an open-loop trajectory error from 3.2% to 5.3% on a 101.4mm long catheter. A two-level motion planning method was used in [17] to decrease the error and discontinuity generated by the iterative Jacobian-based approach. It achieves an error between 0.5 and 5.6mm on a 12-section CR with a length of 1235mm and a diameter of 45 mm. Owing to the non-linear errors and the non-convex nature of the solution set for the inverse kinematics (IK) of multi-section continuum robots (CRs), the model-based IK approaches often result in inaccuracies and entail complex, extensive calculations.

To address this issue of inverse kinematics, several learning-based inverse kinematic algorithms have been investigated to avoid the complex models of continuum robots and learn the nonlinearity. For example, a multi-layer perceptron (MLP) was used to map the workspace to the actuation space for a two-section CR [18] and achieved a simulation error of 2.90% of the total robot length. However, this approach is only suitable for the CRs with a convex IK solution set, as the mapping from the task space to actuation space for CRs with multiple sections (more than two) can be non-convex. The non-convex problem was solved in [19] by using the shape of the continuum robot at the previous step and the desired position to determine the shape at the next step. This method achieved a total robot length error of 3.5% on an 850mm long three-section CR. The weakness of learning-based IK is its data dependency. The inherent uncertainty and low precision of CRs make generating noise-free, relevant training data a formidable task. In addition, the application of online learning methods, such as reinforcement learning [9][20], has also been studied. A multi-agent deep Q network was used to control a single-section, 30 mm flexible CR [9]. It achieved an average error of about 1% of the robot length. However, this method entails significant random movements in the initial learning phases, which may pose risks for continuum robots [21].

Unlike traditional industrial robots, which can effectively operate with open-loop control due to their high accuracy, continuum robots (CRs) typically exhibit lower precision in open-loop scenarios. This necessitates the adoption of closed-loop control systems to achieve higher accuracy in applications such as welding and coating fields [22]. Various methods have been proposed to address the strongly coupled and highly non-linear challenges associated with CRs. Similar to the modeling of CRs, the close-loop control is divided into model-based and model-free methods. Some approaches use existing kinematic models to linearize the CR and then control it with a linear controller. Feedback linearization was used for the closed-loop control of a two-section pneumatic soft continuum robot and achieved a tip error in a range of $\pm$4mm [23]. A puller-follower controller was designed to control a 12-section cable-driven continuum robot with an average tracking error of 0.385% of the robot length [7]. These methods depend on the model's accuracy and the adequate compensation for non-linear errors. In contrast, several learning-based control methods, including neural networks and reinforcement learning, have been employed to circumvent the need for explicit modeling and the design of non-linear controllers. Three neural networks were used to achieve closed-loop control for a 6 DoFs CR with a tip positioning error of 2mm [24]. Further, deep reinforcement learning was used to control a two-section, 200mm long pneumatic CR [25]. The average error was 5.9%-7.6% for different paths.

The above-mentioned learning-based methods have been successfully demonstrated on many CRs. Still, closed-loop control of continuum robots with better positioning and tracking accuracies (e.g., less than 1mm max errors) remains challenging. The challenge is predominantly from its various non-linear factors and the coupling caused by its design structure—the non-linear errors, such as cable elongation and friction between the cable and the vertebrae. So, model-based control is usually difficult to design and requires complex computation. The learning-based control method can learn this uncertainty to a certain extent, but there are difficulties in data collection and training processes. In addition, a well-trained neural network is unsuitable for different robots, meaning every distinct CR design requires a tailored training process. Learning-based methods are also poorly interpretable. For example, in some applications with high safety requirements like surgery, the limited interpretability can pose safety concerns, as the robots may act unexpectedly.

This paper investigated a model-less closed-loop controller based on fuzzy logic for a three-section cable-driven continuum robot. Compared to those model-based methods, this fuzzy controller is more general and insensitive to the model mismatch of continuum robots. It is fully interpretable and does not require a strenuous learning process. Further, it can solve the highly coupled, non-linear, and highly uncertain close-loop control problem of the CRs. Section II will discuss the challenges of using fuzzy logic controllers. Section III will introduce the design process of the fuzzy logic controller, and Section IV will give experimental results to prove the effectiveness and robustness of the controller.

## II. Challenges

Many efforts have been made on the close-loop control of continuum robots. In CRs with multiple sections (expressly, those encompassing more than two sections), multiple inverse kinematics solutions necessitate acquiring the robot's shape information for feedback control. However, achieving accurate shape reconstruction is challenging in practice. Factors like the inherent uncertainty of the CR, including elastic deformations, cable elongation, and external disturbance, reduce the accuracy of model-based shape reconstruction. An alternative solution is to leverage sensors equipped with shape detection capabilities, including vision-based sensors and fiber Bragg grating (FBG) sensors. While these methods provide more accurate shape feedback, they also significantly increase computational demands.

Fig. 1(a) illustrates a typical three-section cable-driven CR. Each section possesses two degrees of freedom, culminating in an overall six degrees of freedom for the robot. Fig.1(b) depicts the transition between the actuation space (representing cable displacement), the configuration space (parameters delineating the CR's shape), and the task space (the Euclidean space) [13]. Due to the non-linear factors, discrepancies arise in mapping between the actuation and configuration spaces. Moreover, the presence of the configuration space leads to the need for additional computations.

In this paper, a new model-less fuzzy logic controller of the continuum robot is proposed. This controller directly maps the task space to the actuation space and is equipped with a robust, built-in shape reconstruction method capable of resolving issues arising from the multiple solutions of inverse kinematics (IK). Contrasting with state-of-the-art approaches, this novel shape reconstruction method requires only the tip position and the sign of the driven cable displacement while maintaining high robustness against external interference. Consequently, only the position and orientation feedback of the robot's tip are necessary to implement closed-loop control using the proposed fuzzy logic controller. This strategy effectively minimizes the required sensors. Therefore, position sensors such as IMUs can be used instead of expensive and complex shape sensors like Fiber Bragg Grating (FBG), reducing costs, computational effort, and space requirements within the continuum robot.

The coupling and nonlinearity of continuum robots significantly affect the formulation of fuzzy rules, which are the cornerstone of fuzzy controllers. Until now, few studies have been on the fuzzy rules' expression for the motion of continuum robots. Relying excessively on the model when establishing these rules can compromise their robustness in the face of uncertainties. The proposed controller's fuzzy rules minimally depend on the robot's mathematical models, making it universal and insensitive to model mismatch. Furthermore, its design is simplified by taking advantage of the coupling properties between different sections, which will be discussed next.

## III. Methodology

In this section, the design process of the proposed fuzzy logic controller for a multi-section cable-driven continuum robot will be introduced. First, a shape reconstruction method that does not require shape sensor and precise calculations of kinematics mode is proposed. Then, the coupling properties and fuzzy rules are discussed. Next, the membership functions of the fuzzy controller are designed to realize fuzzification and defuzzification. Finally, the initial state and singularity of the proposed controller are analyzed.

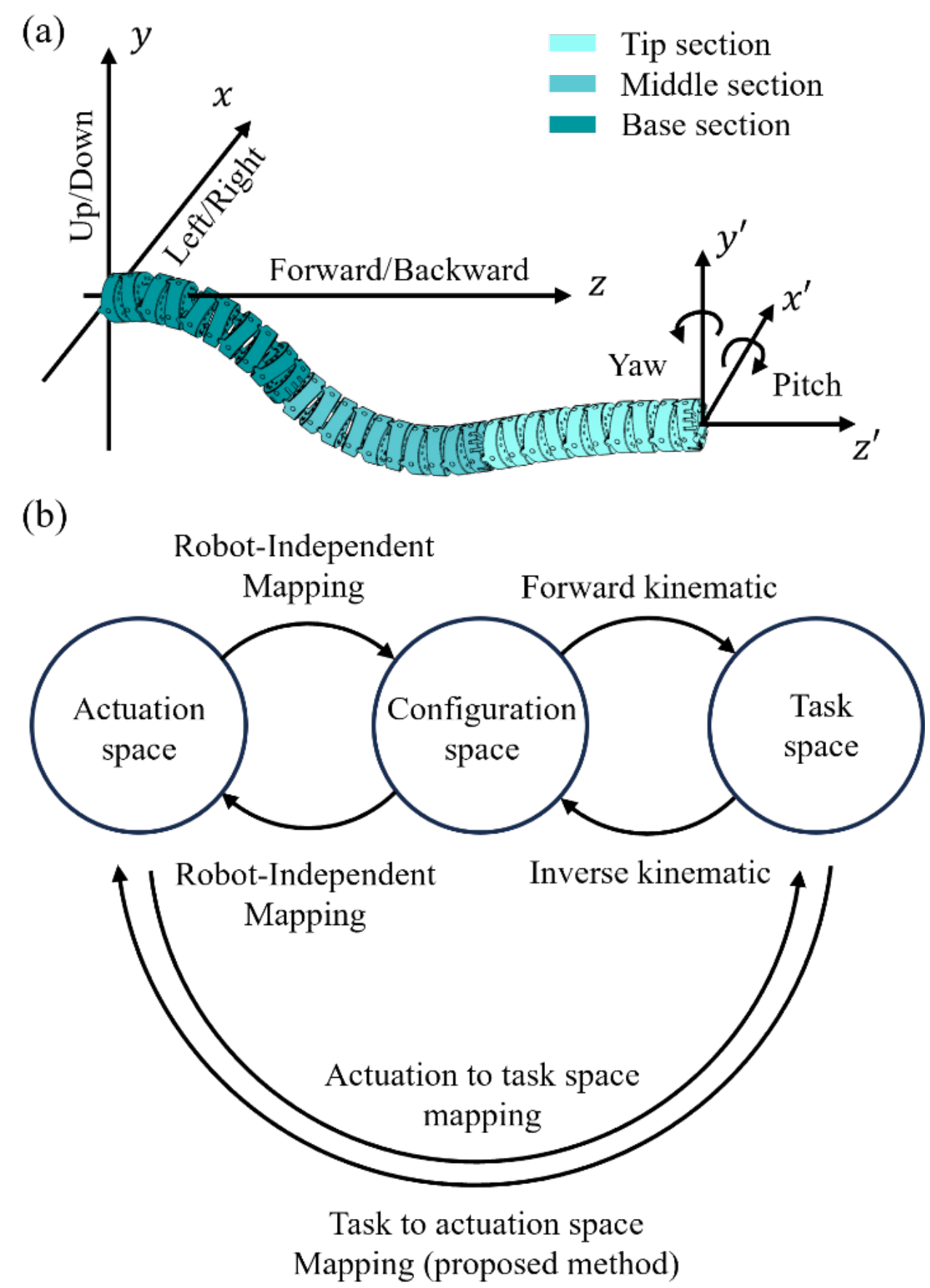

**Fig. 1.** (a) A three-section continuum robot. (b) The mapping between actuation, configuration, and task space [13].

### A. System structure

In Fig.2, a top view of a general vertebra design of a cable-driven continuum robot is depicted. $l_i'$ are the actual drive cables. Typically, angles exist between the drive cables of different sections. To make the fuzzy rules more concise, ideal driving cables $l_i$ are used in fuzzy controller calculations. The ideal driving cables are aligned with the X and Y-axes in the world coordinate system, respectively.

Take the base section (see Fig.1(a)) as an example, in Fig.2, the actual driving cables of base section are $l_1' \sim l_4'$. The angle between the ideal driving cables and the actual driving cable is $\alpha$. The transition from the ideal driving cables $l_i$ to the actual driving cables $l_i'$ for the base section is represented by the equations (1), (2), (3), and (4).

$$\varphi_{base} = tan^{-1}\left(\frac{l_4 - l_2}{l_3 - l_1}\right) \tag{1}$$

$$\theta_{base} = \frac{l_1 - 3l_2 + l_3 + l_4\sqrt{(l_4 - l_2)^2 + (l_3 - l_1)^2}}{d(l_1 + l_2 + l_3 + l_4)(l_4 - l_2)} l_{base} \tag{2}$$

Where $l_i, i = 1,2,3,4$ is the cable length. $l_{base}$ is the length of the base section. $\theta_{base}$ and $\varphi_{base}$ are the bending angle and the direction angle of the base section. d is the radius of the robot's vertebra. As shown in Fig. 2, the four cable holes in the base section need to be rotated by $\alpha$ degrees clockwise to align with the X and Y axes. The actual driving cable length $l_i'$ can be obtained from equations (3) and (4).

$$\begin{cases} l_i' = l_{base} - \theta d cos\varphi_i' \\ \quad i = 1,2,3,4 \end{cases} \tag{3}$$

Where

$$\varphi_i' = \varphi_{base} + \frac{\pi(i-1)}{2} - \varphi_{offset} \tag{4}$$

$\varphi_{offset}$ is the offset angle between the cable hole and the X-axis. For the base section, $\varphi_{offset} = \alpha$. $\theta d cos\varphi_i'$ is the cable displacement after the transformation.

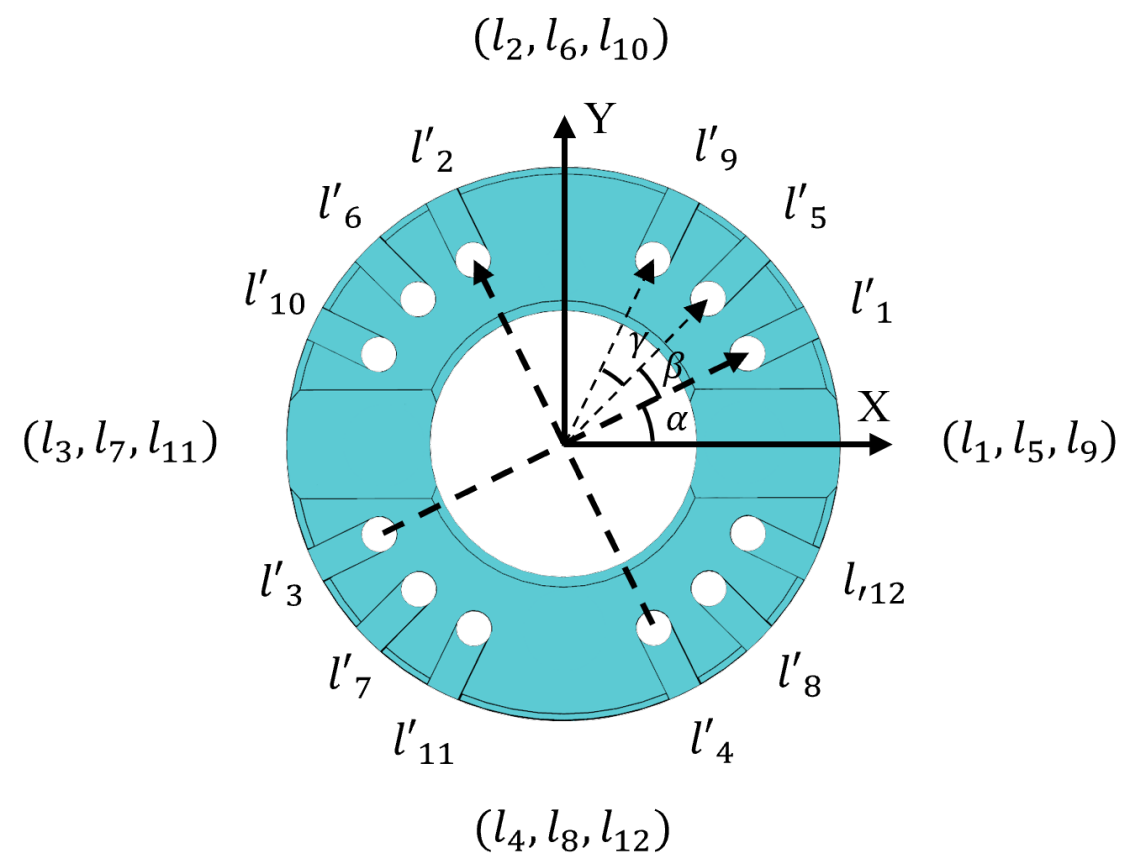

**Fig. 2** Top view of a general continuum robot's vertebrae. $l_1 \sim l_4$, $l_5 \sim l_8$ and $l_9 \sim l_{12}$ are the ideal driving cable of base section, middle section and tip section respectively.

It can be found from Fig. 2 and Eq. (4) that $\varphi_3' - \varphi_1' = 180°$. According to Eq. (3):

$$\Delta l_1' = -\theta d cos\varphi_1' = \theta d cos\varphi_3' = -\Delta l_3' \tag{5}$$

In the same way, $\Delta l_1 = -\Delta l_3$. When $l_1, l_2, l_5, l_6, l_9, l_{10}$ are obtained, the remaining cable length can be calculated using $\Delta l_1 = -\Delta l_3$, $\Delta l_2 = -\Delta l_4$, $\Delta l_5 = -\Delta l_7$, $\Delta l_6 = -\Delta l_8$, $\Delta l_9 = -\Delta l_{11}$, $\Delta l_{10} = -\Delta l_{12}$. While the friction and cable elongation might make the above formulas slightly imprecise, the proposed fuzzy controller is adept at accommodating these variations. The experiments demonstrated that doing so did not significantly impact the control performance.

As shown in Fig. 3(a), the system comprises a fuzzy logic controller, a three-section CR illustrated in Fig. 1(a), and a feedback system for position and orientation. The system input is defined by the desired position $(x_d, y_d, z_d)$ and the desired orientation represented by Euler angle Pitch and Yaw $(P_d, Y_d)$. Roll is ignored as it represents the twist around the Z-axis, a parasitic motion that cannot be controlled independently in the above robots. The system's output is the actual position $(x_a, y_a, z_a)$ and orientation $(P_a, Y_a)$ of the robot's end.

Fig. 3(b) presents the input and output of the proposed fuzzy logic controller. This controller focuses on changes in cable length, bypassing intermediate variables in the configuration space and simplifying the system's complexity.

*B. Shape reconstruction*

When multiple solutions exist in the inverse kinematics, as shown in Fig. 4, it becomes essential to reconstruct the CR's shape for closed-loop control. However, when using model-based methods to reconstruct CR's shape based on drive cable displacements, the accuracy can be compromised by factors such as cable elongation, elastic deformation, and friction. Incorrect shape information will potentially cause significant instability in the feedback control system.

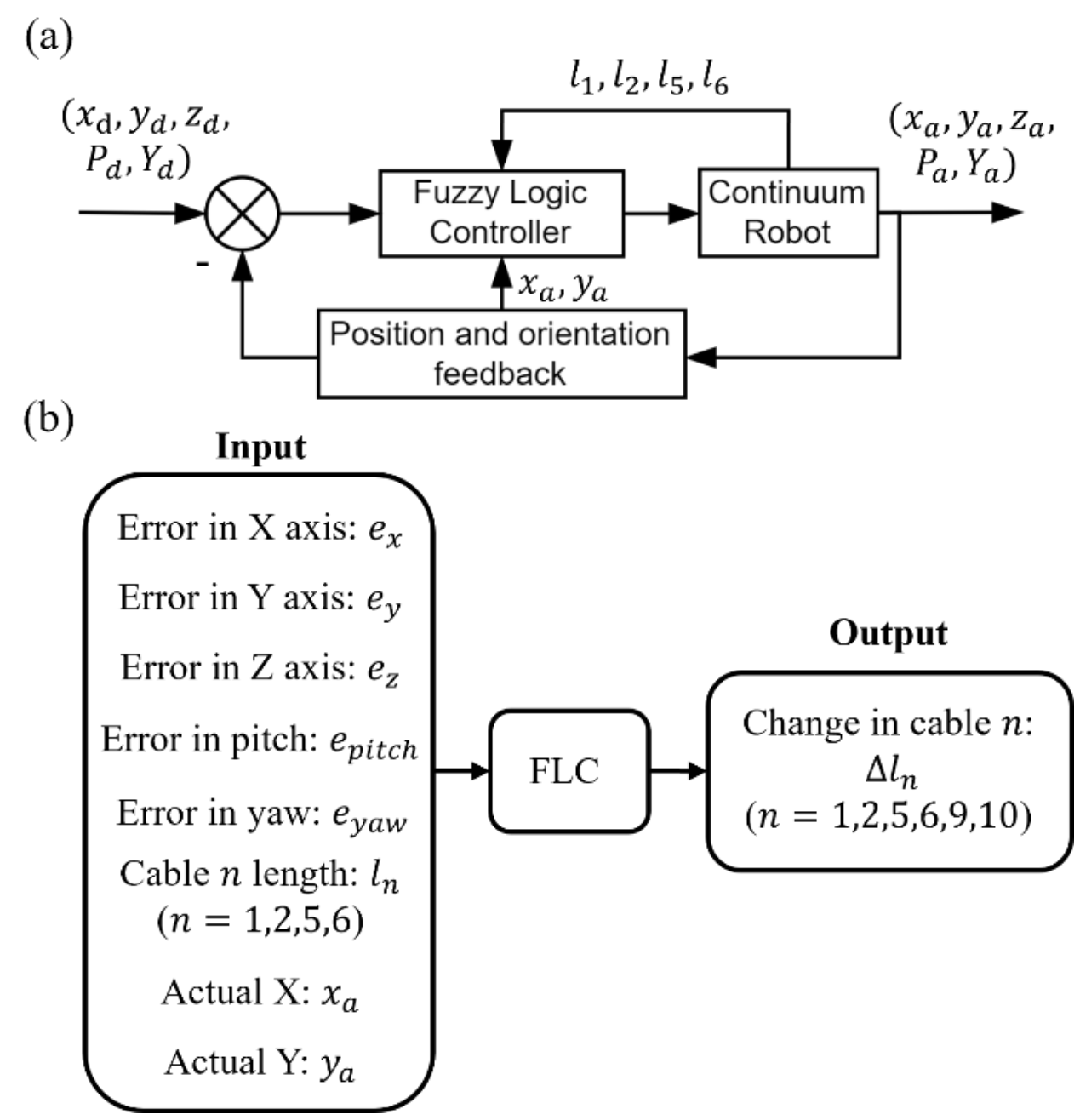

**Fig. 3** (a) System structure. (b) The input and output of the fuzzy logic controller.

Here, a fast, model-less approach that enables fuzzy controllers to sense the robot's shape is proposed. This approach cannot obtain precise shape information, as an accurate reconstruction of the continuum robot's shape is complex, and only position and orientation feedback are needed. This study found that the shapes of robots can be divided into four categories by the phase angle $\varphi_m$ in the middle section. According to (1), $\varphi_{mid}$ can be written as

$$\varphi_{mid} = tan^{-1}\left(\frac{l_8 - l_6}{l_7 - l_5}\right) \tag{6}$$

As $\Delta l_7$=−$\Delta l_5$, $\Delta l_8$=−$\Delta l_6$, Eq. (6) can be rewritten as

$$\varphi_{mid} = tan^{-1}\left(\frac{\Delta l_6}{\Delta l_5}\right) \tag{7}$$

The pseudo-code of the algorithm is as follows:

**Algorithm 1** Shape reconstruction

$\varphi_{mid} = tan^{-1}\left(\frac{\Delta l_6}{\Delta l_5}\right)$
**if** $(l_7 - l_5) < 0$
  $\varphi_{mid} = \varphi_{mid} + 180^0$
**end if**
**if** $\varphi_{mid} < 0$
  $\varphi_{mid} = \varphi_{mid} + 360^0$
**end if**

Where $\varphi_{mid} \in [0, 360]$. However, the exact value of the phase angle $\varphi_{mid}$ does not need to be known. According to Algorithm 1, when the sign of $\Delta l_5$ and $\Delta l_6$ are known, the range of $\varphi_{mid}$ can be obtained. Table II illustrates the correlation between the sign of cable displacement in the middle section and the respective shape categories.

TABLE I
THE RELATIONSHIP BETWEEN THE CABLE LENGTH OF THE MIDDLE SECTION AND SHAPE CATEGORIES

| $\Delta l_5$ | $\Delta l_6$ | $\varphi_{mid}$ | Categories |
|---|---|---|---|
| $< 0$ | $\leq 0$ | $\left[0, \frac{\pi}{2}\right)$ | I |
| $\geq 0$ | $< 0$ | $\left[\frac{\pi}{2}, \pi\right)$ | II |
| $> 0$ | $\geq 0$ | $\left[\pi, \frac{3\pi}{2}\right)$ | III |
| $\leq 0$ | $> 0$ | $\left[\frac{3\pi}{2}, 2\pi\right)$ | IV |

As shown in Fig. 4, two robots have the same base and tip positions but in different shapes: configuration one and two. Their colors, from the base section to the tip section, correspond to a gradient from dark to light. From the top view, it can be observed that the phase angle of the middle section of the cyan-blue robot (configuration one) falls within the range of $\left[0, \frac{\pi}{2}\right)$. So, it belongs to categories I ($\varphi_{mid} \in \left[0, \frac{\pi}{2}\right)$). In contrast, the orange one (configuration two) is in categories IV ($\varphi_{mid} \in \left[\frac{3\pi}{2}, 2\pi\right)$). It is worth noting that even in the same category, there can still be more than one inverse kinematics solution. However, robots under the same category share the same fuzzy rules, which will be discussed later.

With the above method, the fuzzy controller can determine the four different shapes of the robot by comparing the sign of the cable's change ($\Delta l_5$ and $\Delta l_6$) in the middle section. This method is immune to hysteresis and cable elongation because it does not require the accurate value of cable displacement. Although there is a potential for uncertainty to result in incorrect sign determination when $\Delta l_5$ or $\Delta l_6$ is near zero, it's important to note that the CR is also at a critical juncture between two distinct categories in these instances.

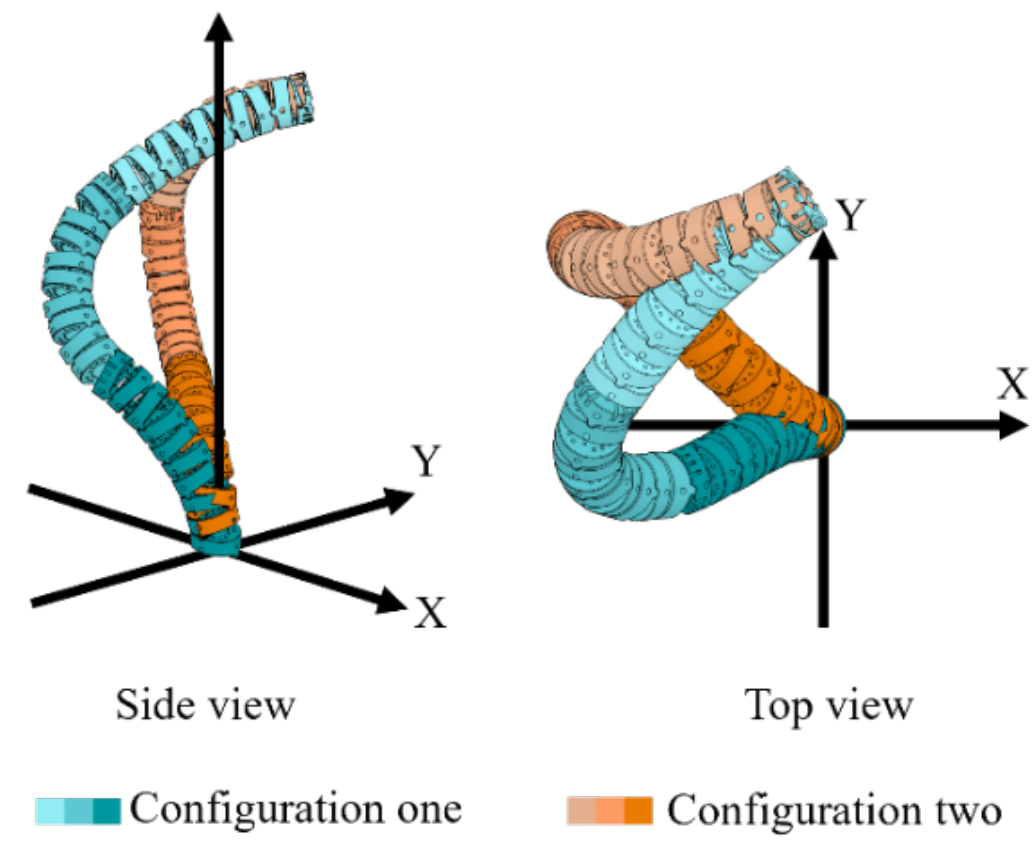

**Fig. 4.** Side and top views of two CRs configurations that reach the same position with the same orientation.

*C. The coupling properties and orientation control*

In multiple sections cable-driven CRs, the position and orientation of the tip is influenced by coupling between different sections of the robot. When the section proximal to the base is actuated, it necessitates an adjustment in the driving cable of the distal section to compensate for the resultant change in cable length caused by the movement of the proximal section. In CRs with three sections, if the driving cable in the tip section is intentionally not compensate for the movement in the base and middle section, the robot's orientation will tend to remain unchanged due to the constraints of cables (see Video 1).

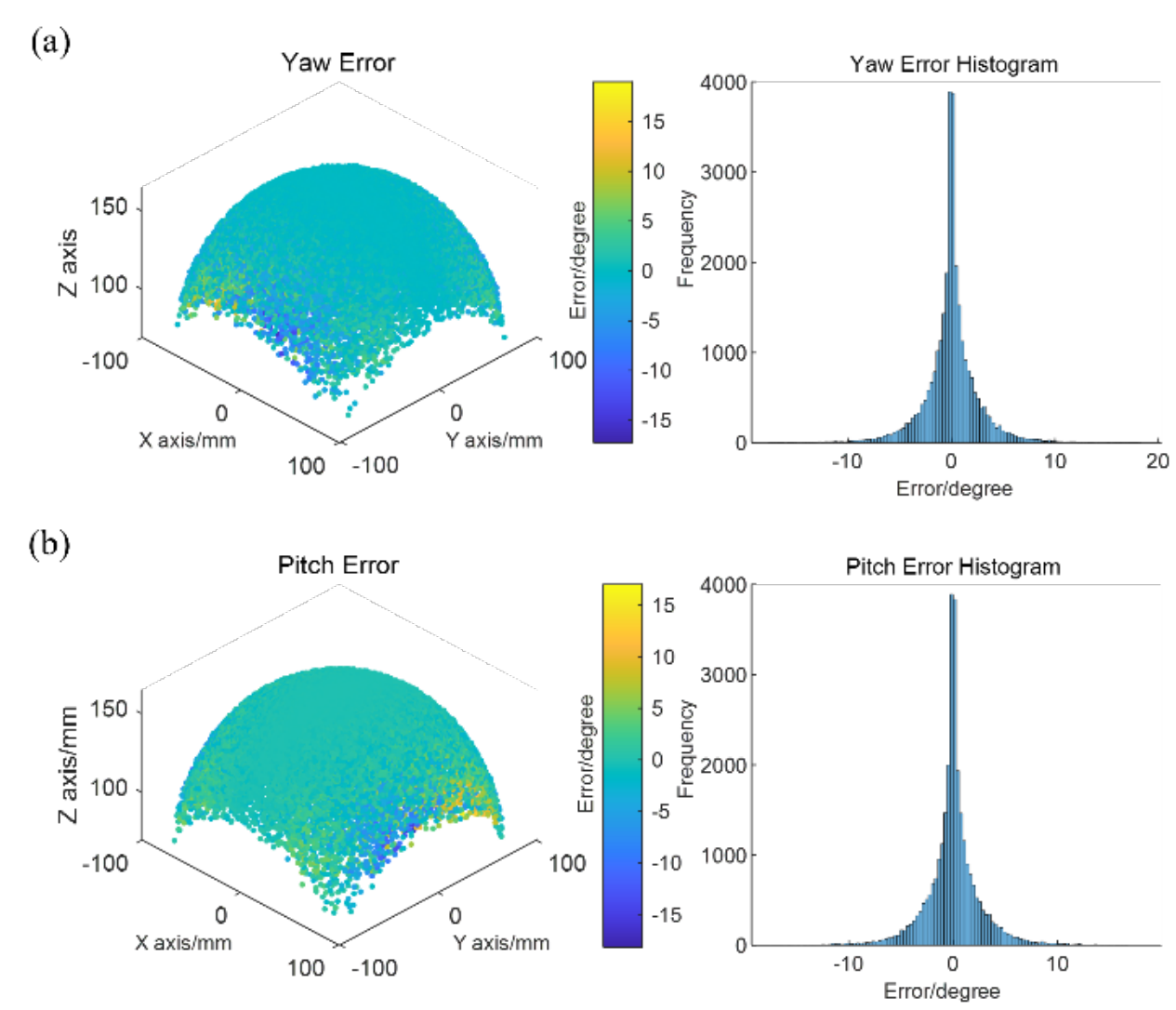

**Fig. 5.** The results of the error simulation, when the tip section does not compensate for the movements of the mid and base sections, include (a) analysis of yaw error across the entire workspace and its corresponding histogram; (b) analysis of pitch error across the entire workspace and its corresponding histogram.

It is essential to acknowledge that the orientation will still experience slight changes in response to the movements of the base and mid sections, even when the driving cable of the tip section is fixed, as shown in Fig. 5. For the yaw error, 97.90% of the data falls within three standard deviations of the mean (±7.91°). Similarly, 98.00% of the data is contained within three standard deviations of the mean (±7.94°) for the pitch error.

Compared with the model-based controller that performs position and orientation control on the entire CR at the same time [7], utilizing the coupling characteristic allows orientation remains nearly unaffected when the position changes. Therefore, position and orientation can be decoupled. The proposed controller uses only the tip section for orientation control. Exploiting these coupling properties significantly simplifies orientation control and the associated fuzzy rules. Moreover, reducing the coupling between position and orientation improves the system's robustness.

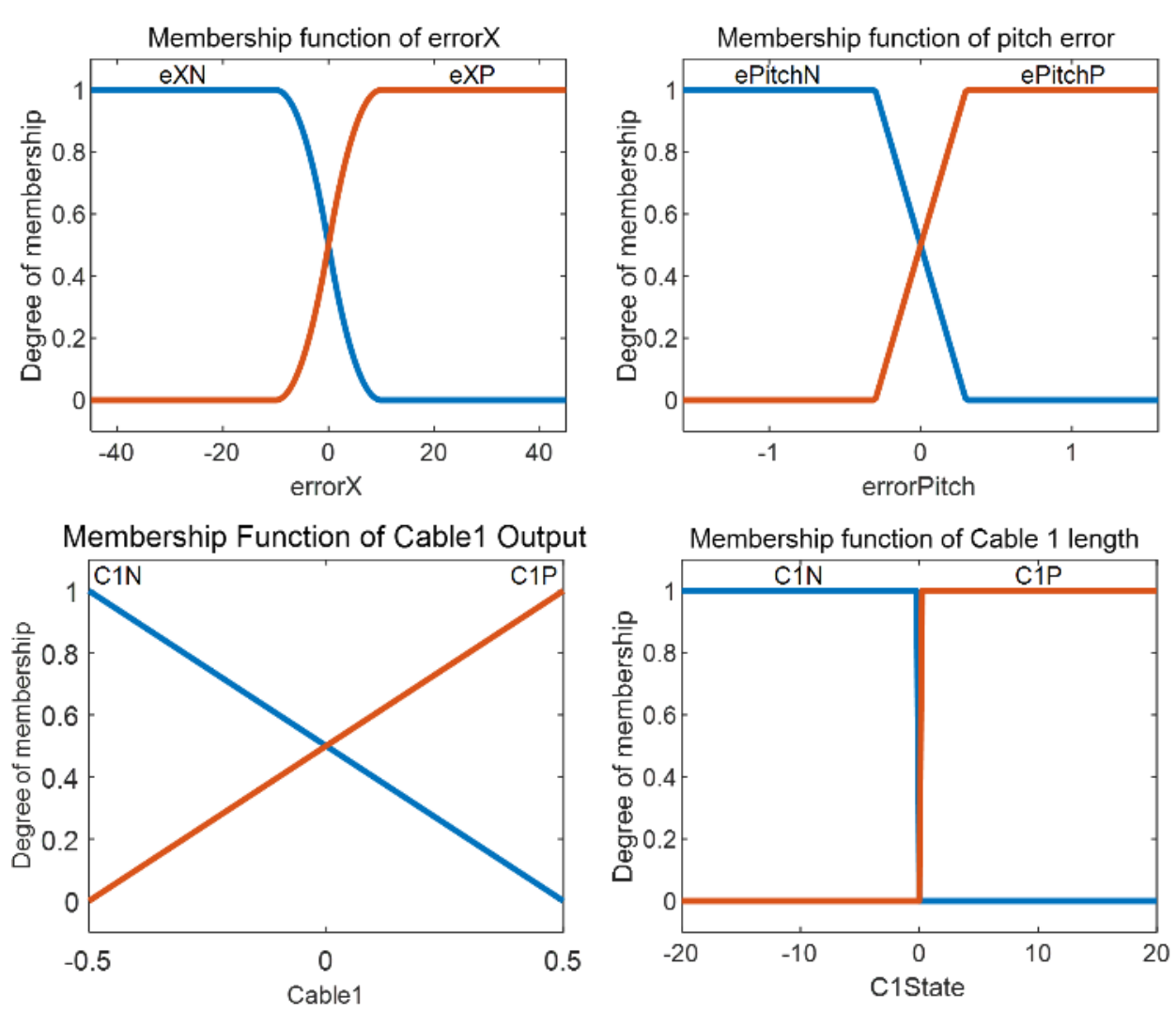

**Fig. 6.** Membership function.

*D. Fuzzy logic controller design*

Here, the design process of the model-less fuzzy logic controller for a three-section continuum robot is introduced. The process of the fuzzy logic controller is divided into three steps: fuzzification, inference, and defuzzification. Some membership functions are shown in Fig. 6. To reduce the number of rules, after multiple tests, each input variable's membership functions are set at two, one Z-shaped membership function (ZMF) and one S-shaped membership function (SMF). These membership functions categorize the input error into two types: positive error ($eP$) and negative error ($eN$), similarly classifying changes in input cable length as either negative ($CN$) or positive ($CP$).

*1). Position control.*

As previously discussed, the approximate shape of the three-section continuum robot can be obtained from $\Delta l_5$ and $\Delta l_6$. Suppose two CRs with different shapes have the same base, tip position, and orientations. Fig. 7(a) shows its X-Z plane.

The cable displacement of configuration one is shown in Fig. 7(b). In motions one and four, $l_7^1$ becomes longer. This is because when the Z coordinate remains unchanged, the decrease in the X and Y coordinates causes the robots to accumulate in the center area of the coordinates. The robot is, therefore, in a crouched position. This aggravates the bending of the middle section. For motions two and three, the robot stretches due to the increase in X and Y coordinates, so the curvature of the middle section decreases. For the base section, the bending direction is the same as the direction of movement. So, for motions one and two, $l_3^1$ will be longer. The tip section is used for orientation control, which will be discussed later.

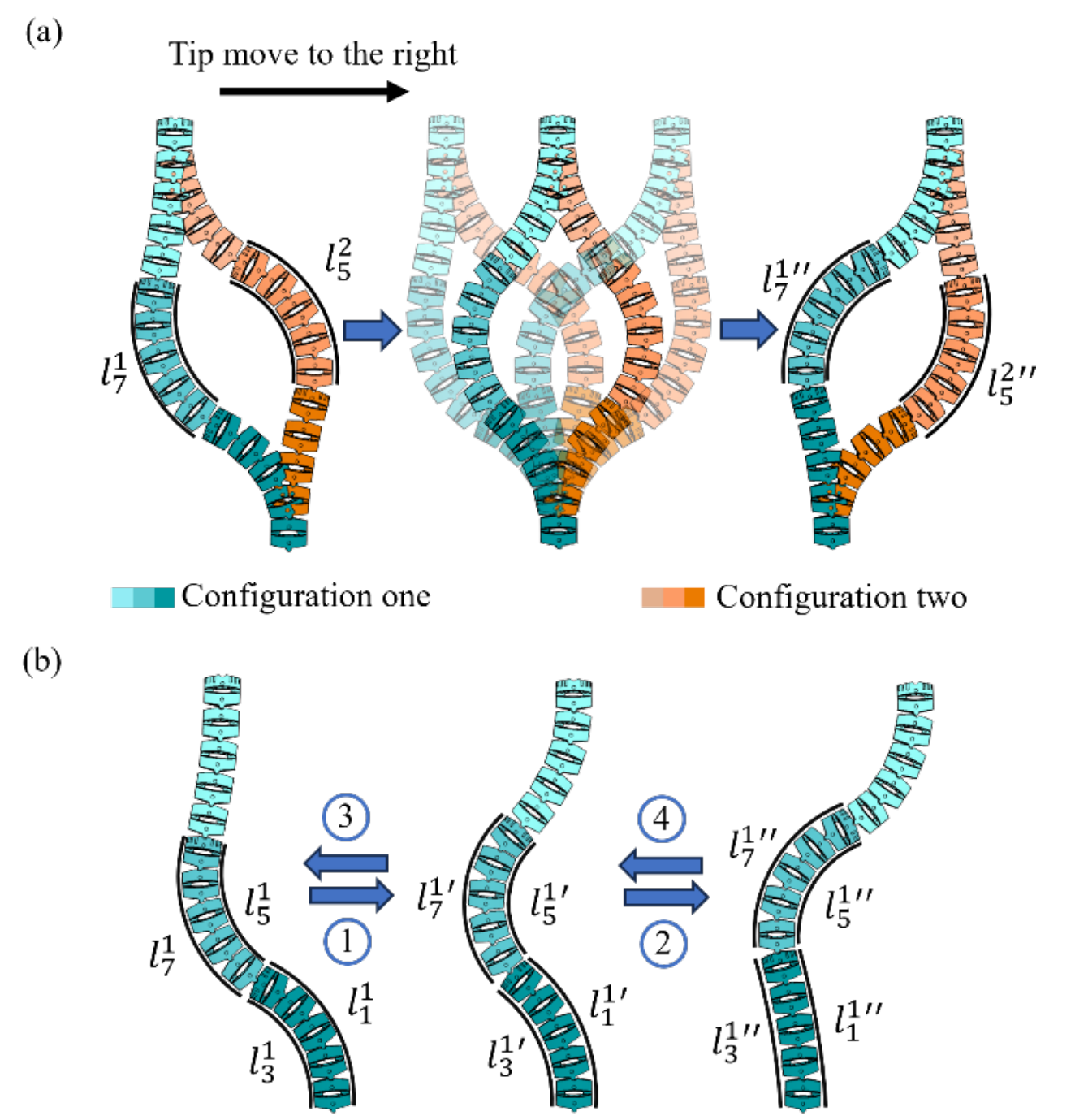

**Fig. 7.** Motion analysis. (a) The translational motion of a CR under different configurations. (b) The analysis of cable displacement. ①, ②, ③, and ④ represent motions one, two, three, and four. Motion one and four represent the movement from left or right to the middle, respectively, while motion three and two represent the movement from middle to left and right.

For configuration two, since the inner and outer sides of the middle section arc are opposite to configuration one, the changing directions of $l_7^1$ are also opposite to those of configuration one. Due to the continuity of the robot's shape changes during the movement, CR will not jump from configuration one to configuration two.

In the Y-Z plane, the same rules apply. Combined with the shape reconstruction in Section III. b, Table II can be obtained.

As for the motion along the Z-axis, it can be found that when the Z-axis coordinate reaches its maximum value, the robot is fully straightened. In this case, the cable displacement can be expressed as

$$\Delta l_1 = \Delta l_2 = \cdots = \Delta l_8 = 0 \qquad (8)$$

Therefore, if the robot needs to move along the positive direction of the Z-axis, the absolute value of $\Delta l_1$, $\Delta l_2$, $\Delta l_5$, $\Delta l_6$ needs to be decreased. The fuzzy rules used to control Z are shown in Table III.

TABLE II
FUZZY RULES FOR THE X AND Y-AXES

| Input | | | | | Output | | | |
|---|---|---|---|---|---|---|---|---|
| $e$ | $\Delta l_5$ | $\Delta l_6$ | $x_a$ | $y_a$ | $\Delta l_1$ | $\Delta l_2$ | $\Delta l_5$ | $\Delta l_6$ |
| $e_xN$ | $N$ | / | $P$ | / | $P$ | / | $N$ | / |
| $e_xN$ | $N$ | / | $N$ | / | $P$ | / | $P$ | / |
| $e_xP$ | $N$ | / | $P$ | / | $N$ | / | $P$ | / |
| $e_xP$ | $N$ | / | $N$ | / | $N$ | / | $N$ | / |
| $e_xN$ | $P$ | / | $P$ | / | $P$ | / | $P$ | / |
| $e_xN$ | $P$ | / | $N$ | / | $P$ | / | $N$ | / |
| $e_xP$ | $P$ | / | $P$ | / | $N$ | / | $N$ | / |
| $e_xP$ | $P$ | / | $N$ | / | $N$ | / | $P$ | / |
| $e_yN$ | / | $N$ | / | $P$ | / | $P$ | / | $N$ |
| $e_yN$ | / | $N$ | / | $N$ | / | $P$ | / | $P$ |
| $e_yP$ | / | $N$ | / | $P$ | / | $N$ | / | $P$ |
| $e_yP$ | / | $N$ | / | $N$ | / | $N$ | / | $N$ |
| $e_yN$ | / | $P$ | / | $P$ | / | $P$ | / | $P$ |
| $e_yN$ | / | $P$ | / | $N$ | / | $P$ | / | $N$ |
| $e_yP$ | / | $P$ | / | $P$ | / | $N$ | / | $N$ |
| $e_yP$ | / | $P$ | / | $N$ | / | $N$ | / | $P$ |

TABLE III
FUZZY RULES FOR THE Z AXIS

| Input | | | | | Output | | | |
|---|---|---|---|---|---|---|---|---|
| e | $\Delta l_1$ | $\Delta l_2$ | $\Delta l_5$ | $\Delta l_6$ | $\Delta l_1$ | $\Delta l_2$ | $\Delta l_5$ | $\Delta l_6$ |
| $e_zP$ | $N$ | / | / | / | $P$ | / | / | / |
| $e_zP$ | / | $N$ | / | / | / | $P$ | / | / |
| $e_zP$ | / | / | $N$ | / | / | / | P | / |
| $e_zP$ | / | / | / | $N$ | / | / | / | $P$ |
| $e_zP$ | P | / | / | / | $N$ | / | / | / |
| $e_zP$ | / | $P$ | / | / | / | $N$ | / | / |
| $e_zP$ | / | / | $P$ | / | / | / | $N$ | / |
| $e_zP$ | / | / | / | $P$ | / | / | / | $N$ |
| $e_zN$ | $N$ | / | / | / | $N$ | / | / | / |
| $e_zN$ | / | $N$ | / | / | / | $N$ | / | / |
| $e_zN$ | / | / | $N$ | / | / | / | $N$ | / |
| $e_zN$ | / | / | / | $N$ | / | / | / | $N$ |
| $e_zN$ | $P$ | / | / | / | $P$ | / | / | / |
| $e_zN$ | / | $P$ | / | / | / | $P$ | / | / |
| $e_zN$ | / | / | $P$ | / | / | / | $P$ | / |
| $e_zN$ | / | / | / | $P$ | / | / | / | $P$ |

*2). Orientation control*

As discussed in Section III.C, the coupling properties facilitate the simplification of orientation control. Two pairs of cables in the tip section complete the orientation control. $l_9$ and $l_{11}$ are used to control the rotation around the Y-axis, and $l_{10}$ and $l_{12}$ are used to control the rotation around the X-axis. The fuzzy rules used to control yaw and pitch are shown in Table IV.

TABLE IV
FUZZY RULES FOR YAW AND PITCH

| Input | | Output | |
|---|---|---|---|
| $e_{yaw}$ | $e_{pitch}$ | $\Delta l_9$ | $\Delta l_{10}$ |
| $P$ | / | $P$ | / |
| N | / | N | / |
| / | $P$ | / | $P$ |
| / | $N$ | / | $N$ |

*3). Analysis of initial state and singularity.*

As mentioned in the previous section, the fuzzy controller perceives the robot's shape by comparing the cable length in the middle section. When the robot is in the initial state, or the bending angle of the middle section is 0, $l_5 = l_6 = l_7 = l_8$. At this point, the system is at the singularity. The middle section of the robot can bend in any direction. To jump out of this singularity, add a tiny value of $\delta_5$ *and* $\delta_6$ to $l_5$ and $l_6$, respectively.

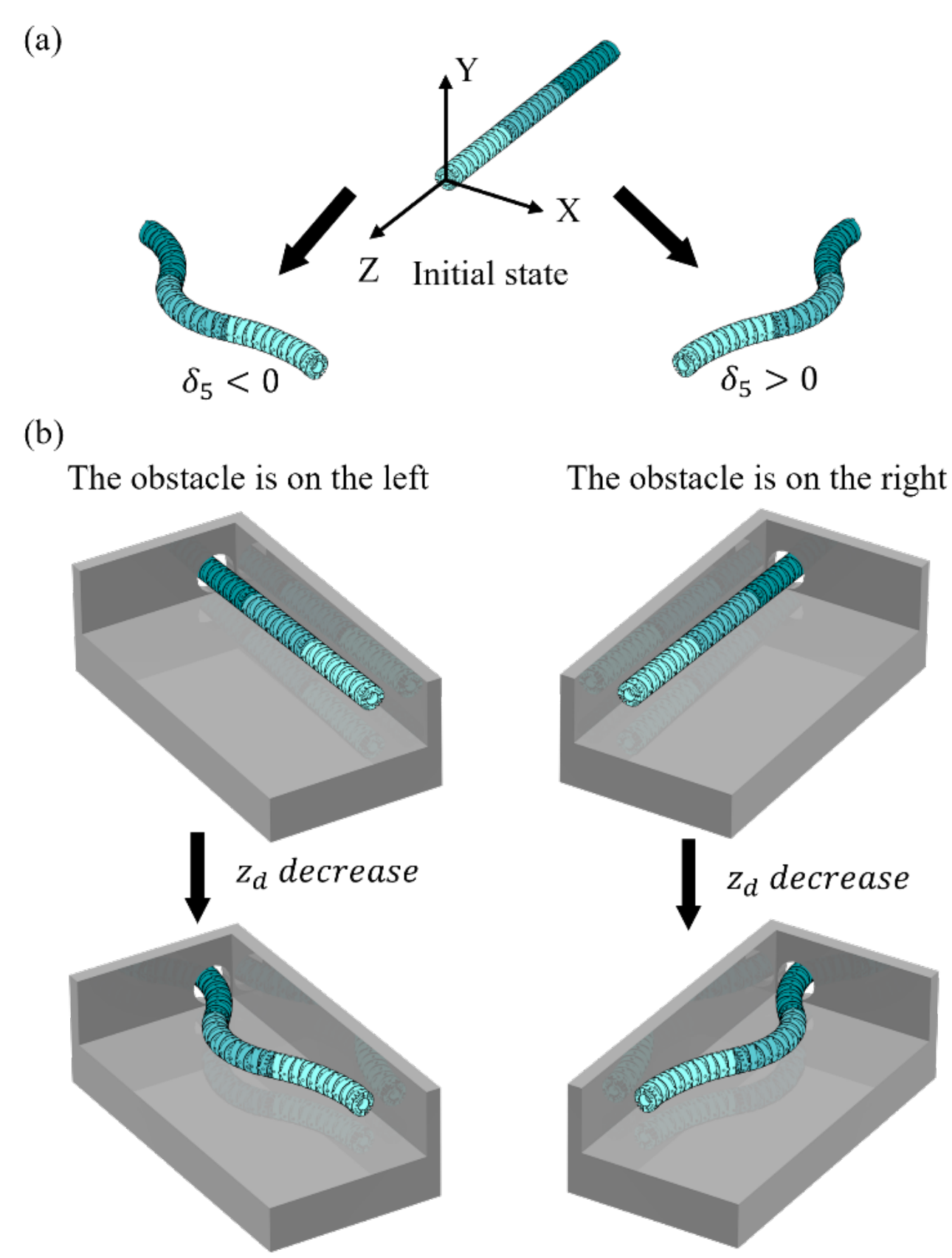

**Fig. 8.** (a) The effect of $\delta_5$ on the robot's shape when the robot is located at a singularity point or initial state. (b) The influence of $\delta_5$ when there are different obstacles in the work area.

When $\delta_5 < 0$ and $\delta_6 < 0$, according to Table I, the robot will be in category one. As shown in Fig. 8(a), different $\delta_5$ result in different shapes for the CR. Therefore, at the singularity point or the initial state, $\delta_5$ and $\delta_6$ can determine the robot's shape for the next period. This feature makes the controller capable of obstacle avoidance, as shown in Fig. 8(b). Likewise, changing between different typical shapes while in motion should be avoided in most cases. $\delta_5$ and $\delta_6$ do not need to be frequently modified.

*4). Fuzzy logic controller*

A Mamdani fuzzy inference system is used in this paper to control the three-section continuum robot. As shown in Fig. 9. The controller is designed with eleven input variables and ten output variables. When it concurrently adjusts errors in the Z-axis and any other axis, the outputs corresponding to these errors may counterbalance each other. Here, $\Delta l_{z1}$, $\Delta l_{z2}$, $\Delta l_{z5}$, and $\Delta l_{z6}$ are the output caused by Z-axis errors. Among them, $\Delta l_{z1}$ and $\Delta l_{z2}$ act on the base section, $\Delta l_{z5}$ and $\Delta l_{z6}$ act on the middle section. The following formulas are used to combine $\Delta l_z$ with $\Delta l$.

$$\begin{cases} \Delta L_1 = k_1 \Delta l_1 + k_2 \Delta l_{z1} \\ \Delta L_2 = k_3 \Delta l_2 + k_4 \Delta l_{z2} \\ \Delta L_5 = k_5 \Delta l_5 + k_6 \Delta l_{z5} \\ \Delta L_6 = k_7 \Delta l_6 + k_8 \Delta l_{z6} \end{cases} \tag{8}$$

$k_1$ to $k_8$ are individual parameters. By adjusting these parameters, the middle section can be more sensitive to the error in the Z-axis, while the base section is more sensitive to the error in the X and Y-axes. In this way, if there is an error in the X, Y, or Z-axis, the controller output will not be zero.

In summary, this chapter has introduced a model-free closed-loop control method utilizing a fuzzy controller. The controller leverages the coupling properties to diminish the coupling between orientation and position, simplifying the control process and enhancing stability. Since it does not rely on mathematical models, this approach effectively circumvents the issues of model mismatch and cable elongation discussed in Section II. The experiments conducted using this controller will be detailed in Section IV.

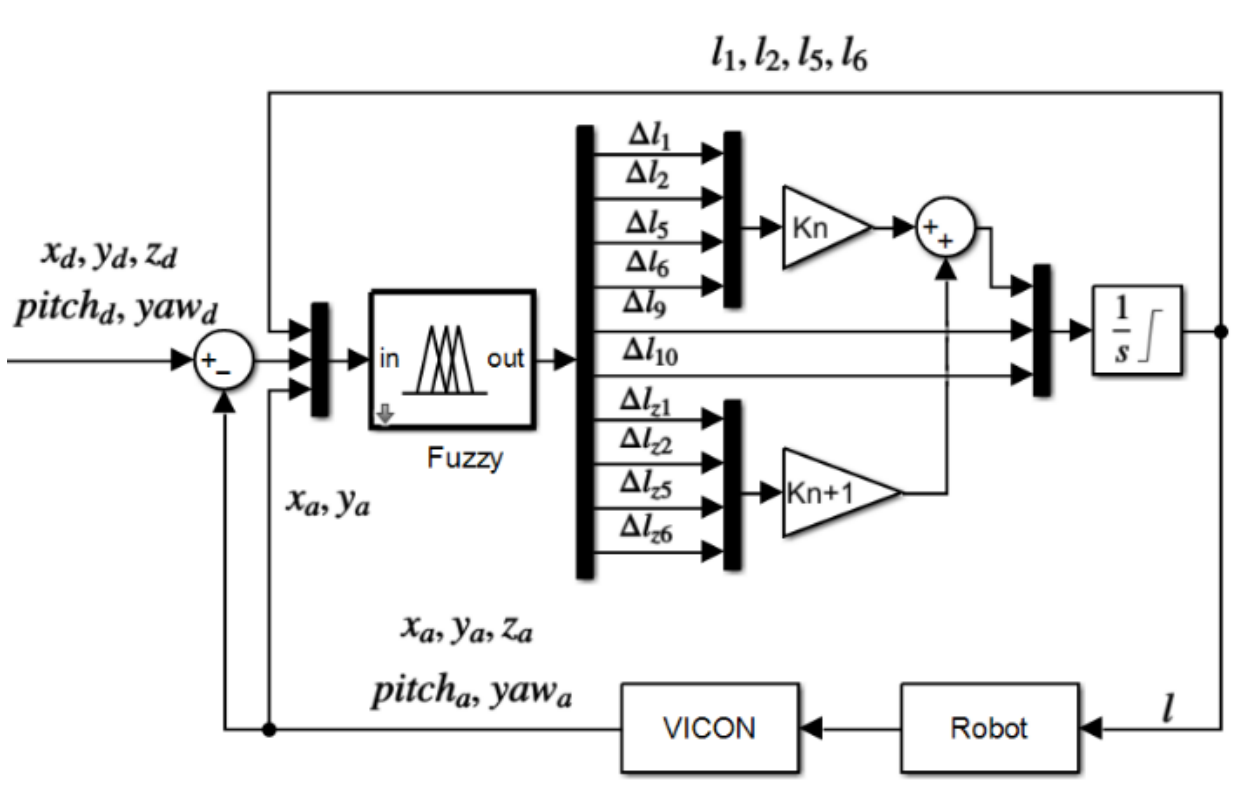

**Fig. 9.** Proposed Mamdani fuzzy inference system (For $K_n$ and $K_{n+1}, n = 1, 2, 3$).

## IV. Experiments

In this section, a set of experiments demonstrating the effectiveness of the proposed fuzzy logic controller is presented. The robot, initially developed in [22], has been modified for closed-loop control. Fig. 10(a) illustrates the experimental setup. A VICON motion capture system is used to provide feedback on the robot's position and orientation for the experiment (0.15mm mean absolute error)[26]. Details of the robot are shown in Fig. 10 (b). The robot comprises three sections, a total length of 165 mm, with an outer diameter of 12.7 mm. Each section offers two DoFs and is actuated by four cables. A ø4mm silicone rod (internal payload simulator) was installed in the working channel of the robot to simulate the cables or hoses of end effectors. Four groups of VICON markers are attached to the robot, located at the end of each section and at the base of the continuum robot. The displacement between the marker and the robot is compensated. The controller parameters are $k_1 = k_3 = k_6 = k_8 = 0.01, k_2 = k_4 = k_7 = k_9 = 0.003$. All the experiments are carried out using two methods, including the proposed fuzzy-logic approach and the primary piecewise contact curves method.

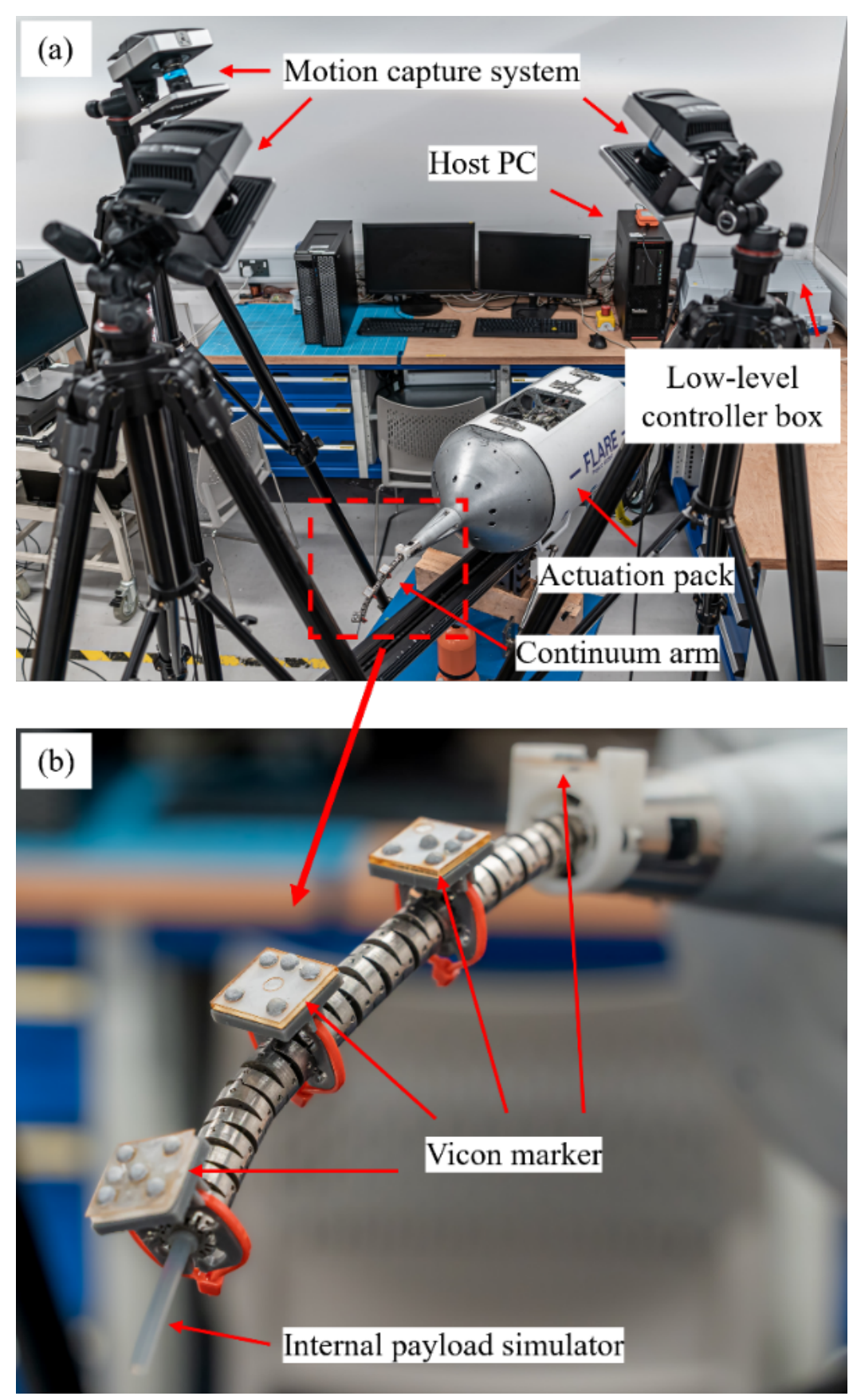

**Fig. 10.** Experimental setup. (a) The motion capture system, host PC, and continuum robot. (b) The details of the Continuum robot being used.

*A. Path follow experiments.*

Comparative experiments assessed the performance differences between the fuzzy logic controller and the PCC (Predictive Control of Constrained) model-based controller. The PCC model-based approach, lacking a shape reconstructor, requires four Vicon markers for individual control of each robot section. In this configuration, control is applied only to the X and Y-axes for each section, as the Z-axis remains passive due to each section's two degrees of freedom (2-DoF) configuration. Consequently, the robot's performance along the Z-axis and its orientation is contingent on the precision of the PCC model. A significant discrepancy in the model can result in considerable errors in these dimensions.

Benefits from the built-in shape reconstruction method, the proposed fuzzy logic requires only two markers to simultaneously control the X, Y, and Z-axes and the robot's orientation—one marker at the robot's tip and another at the base.

Fig. 11 illustrates the path-following results for circular and dual loop-shaped paths. During the path-following process, the desired Euler angles for the orientation of the continuum robot were maintained at zero. These results indicate that the fuzzy controller effectively follows the desired path while maintaining the required orientation, showing a marked improvement over the PCC-based method. The Root Mean Square Error (RMSE) of the path-following error is detailed in Table V.

TABLE V

RMS OF THE PATH TRACKING

| Path | RMSE (mm) |
|---|---|
| Circle (PCC) | 7.37 |
| Circle (FLC) | 0.28 |
| Dual loop-shaped path (PCC) | 7.86 |
| Dual loop-shaped path (FLC) | 0.54 |

### *B. External disturbance experiments.*

To validate the robustness of the proposed model-free fuzzy controller, external disturbances were introduced during the path-following task. Fig. 12 showcases the robustness testing conducted during complex path-following exercises. The results show that the system quickly recovers to become stable when external disturbances are encountered while tracking complex trajectories. Additionally, the external disturbance test was implemented using the PCC-based method. In all tests, the system produced large oscillations and diverged.

### *C. Summary.*

The experimental results demonstrate the effectiveness and robustness of the fuzzy controller proposed in this paper. The controller successfully achieved a path tracking RMSE of 0.28mm for a circle and 0.54mm for a dual loop shape path, which outperformed the results of the classic PCC-based method (circle RMSE – 7.37mm; dual loop-shape path RMSE – 7.86mm). Further, The experimental results showed that the fuzzy logic controller resists an external instantaneous impact of 100g weights. This indicates that the proposed fuzzy controller is resilient to large external disturbances and is not adversely affected by model mismatches. At the same time, the PCC-based method failed all external disturbance tests.

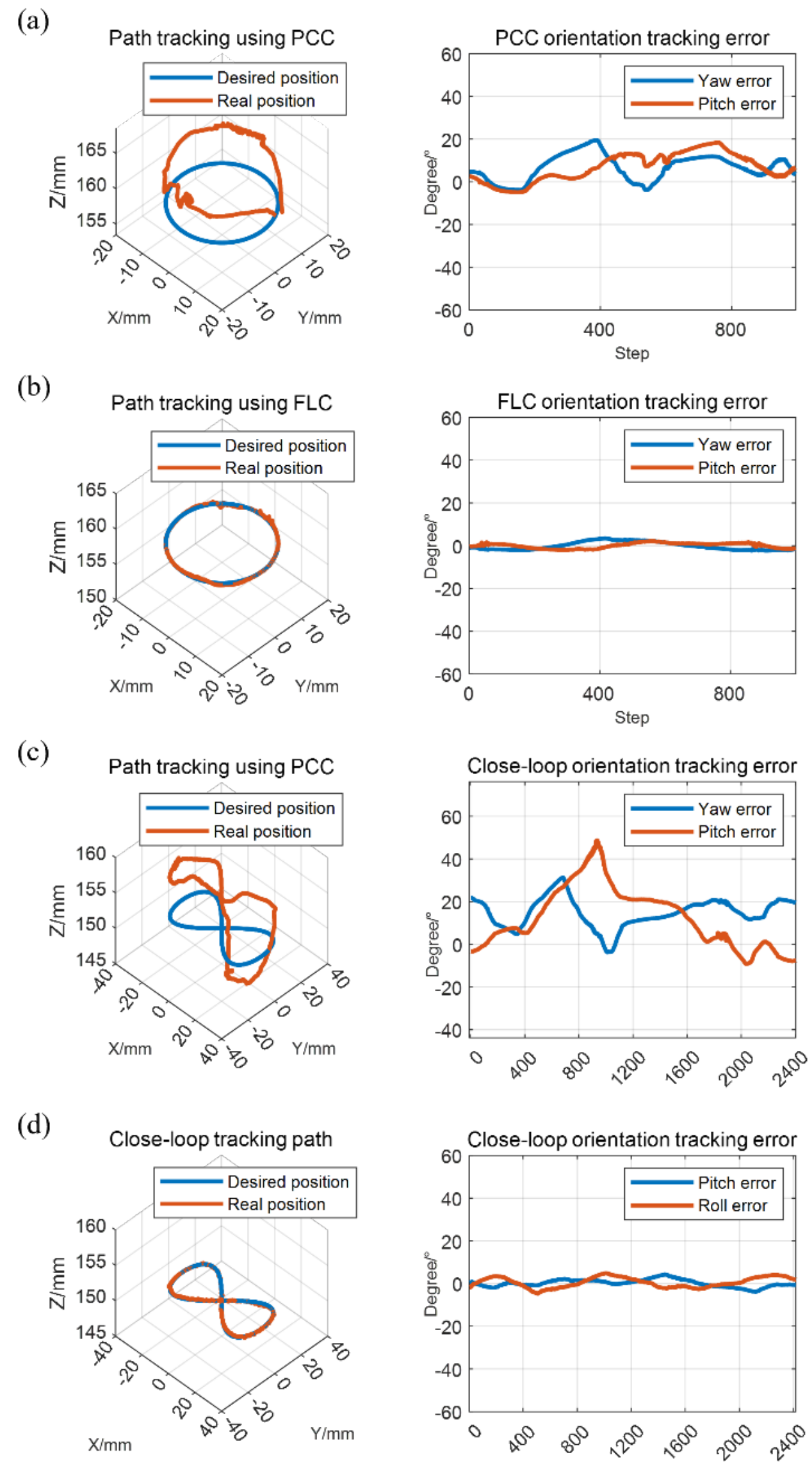

**Fig. 11.** Experiment result (a) Circle path following using the PCC-based method. (b) Circle path following using FLC. (c) Dual loop-shaped path following using PCC-based method. (d) Dual loop-shaped path following.

## V. CONCLUSION

In controlling multi-section continuum robots, factors such as substantial model mismatches, low repeatability, singularities, and motion discontinuities in kinematics significantly impact control performance. Designing an accurate model-based closed-loop control system is typically complex and challenging. Learning-based controllers, on the other hand, require extensive data collection and individualized training for each robot.

This paper introduced a model-less fuzzy logic controller that circumvents the model mismatch issues and does not necessitate prior training. Incorporating a built-in shape reconstructor enables effective closed-loop control of both position and orientation using only the tip's position and orientation feedback, thus significantly reducing reliance on sensors. Utilizing coupling properties simplifies the controller structure and improved system robustness. It also exhibits robust resistance to model mismatch because the fuzzy rules do not rely on mathematical models. This approach allows the proposed controller to attain a RMSE trajectory tracking accuracy of 0.28 mm in a continuum robot with six degrees of freedom and three sections.

In the future, we plan to upgrade the system and implement closed-loop control with access to long passive sections, which will introduce large non-linear errors. Also, we planned to integrate sensors within the CR to facilitate closed-loop control in confined and enclosed environments.

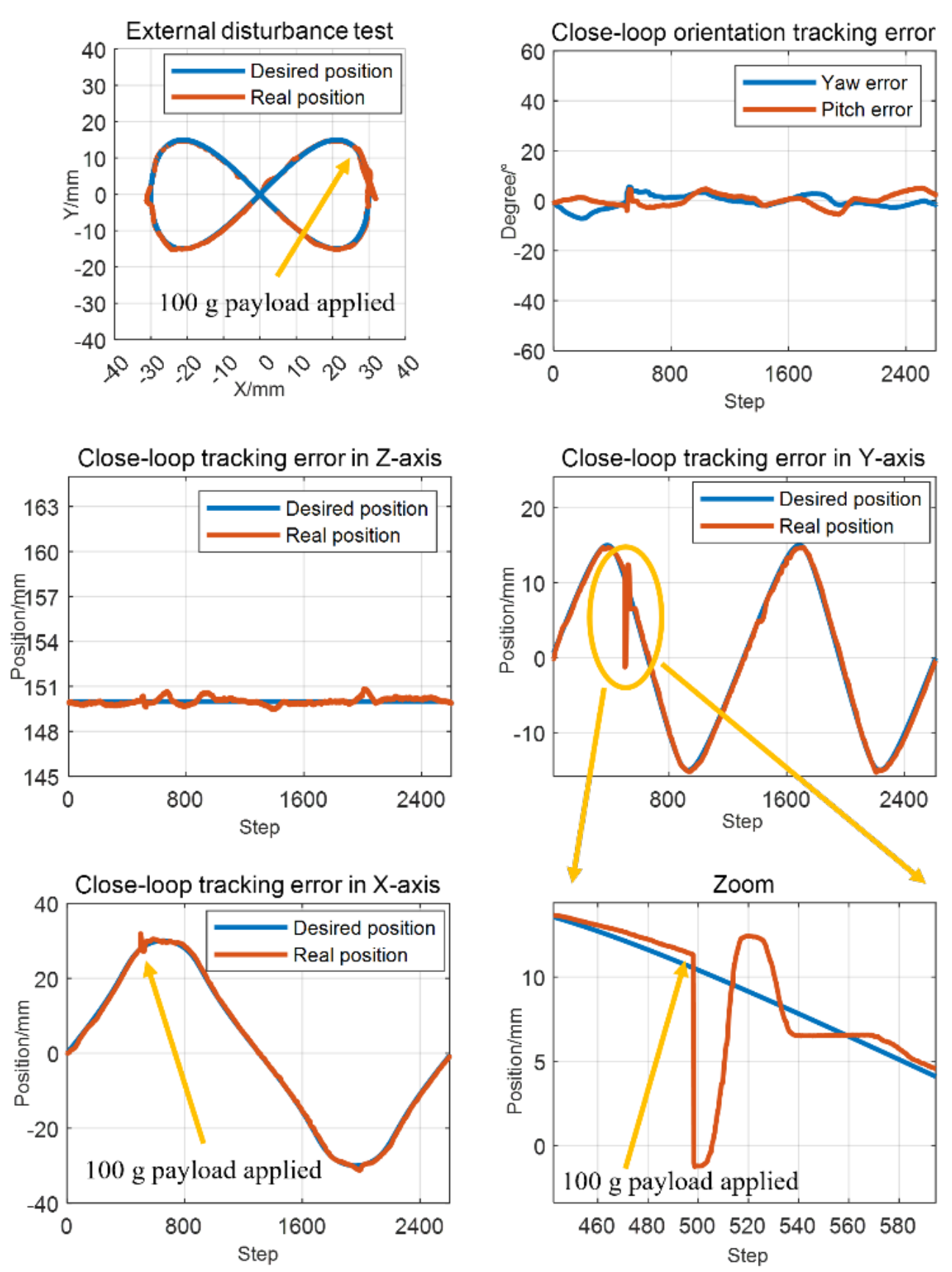

**Fig. 12.** The external disturbance test result for dual loop shape path.

## References

[1] X. Dong *et al.*, "Development of a slender continuum robotic system for on-wing inspection/repair of gas turbine engines," *Robot. Comput. Integr. Manuf.*, vol. 44, pp. 218–229, Apr. 2017, doi: 10.1016/j.rcim.2016.09.004.

[2] Y. Li *et al.*, "An underwater near-infrared spectral continuum robot as a tool for in situ detection and classification," *Meas. J. Int. Meas. Confed.*, vol. 216, p. 112913, Jul. 2023, doi: 10.1016/j.measurement.2023.112913.

[3] D. V. A. Nguyen *et al.*, "A Hybrid Concentric Tube Robot for Cholesteatoma Laser Surgery," *IEEE Robot. Autom. Lett.*, vol. 7, no. 1, pp. 462–469, Jan. 2022, doi: 10.1109/LRA.2021.3128685.

[4] J. Lai, K. Huang, B. Lu, Q. Zhao, and H. Chu, "Verticalized-Tip Trajectory Tracking of a 3D-Printable Soft Continuum Robot: Enabling Surgical Blood Suction Automation," *IEEE/ASME Trans. Mechatronics*, vol. 27, no. 3, pp. 1545–1556, Jun. 2022, doi: 10.1109/TMECH.2021.3090838.

[5] X. Dong, D. Palmer, D. Axinte, and J. Kell, "In-situ repair/maintenance with a continuum robotic machine tool in confined space," *J. Manuf. Process.*, vol. 38, pp. 313–318, Feb. 2019, doi: 10.1016/j.jmapro.2019.01.024.

[6] F. Janabi-Sharifi, A. Jalali, and I. D. Walker, "Cosserat Rod-Based Dynamic Modeling of Tendon-Driven Continuum Robots: A Tutorial," *IEEE Access*, vol. 9, pp. 68703–68719, Jan. 2021, doi: 10.1109/ACCESS.2021.3077186.

[7] Y. Zheng *et al.*, "Design and validation of cable-driven hyper-redundant manipulator with a closed-loop puller-follower controller," *Mechatronics*, vol. 78, p. 102605, Oct. 2021, doi: 10.1016/j.mechatronics.2021.102605.

[8] R. Grassmann, V. Modes, and J. Burgner-Kahrs, "Learning the Forward and Inverse Kinematics of a 6-DOF Concentric Tube Continuum Robot in SE(3)," *IEEE Int. Conf. Intell. Robot. Syst.*, no. 3, pp. 5125–5132, 2018, doi: 10.1109/IROS.2018.8594451.

[9] G. Ji *et al.*, "Towards Safe Control of Continuum Manipulator Using Shielded Multiagent Reinforcement Learning," *IEEE Robot. Autom. Lett.*, vol. 6, no. 4, pp. 7461–7468, Oct. 2021, doi: 10.1109/LRA.2021.3097660.

[10] T. G. Thuruthel, E. Falotico, F. Renda, and C. Laschi, "Model-Based Reinforcement Learning for Closed-Loop Dynamic Control of Soft Robotic Manipulators," *IEEE Trans. Robot.*, vol. 35, no. 1, pp. 127–134, Feb. 2019, doi: 10.1109/TRO.2018.2878318.

[11] J. D. L. Ho *et al.*, "Localized online learning-based control of a soft redundant manipulator under variable loading," *Adv. Robot.*, vol. 32, no. 21, pp. 1168–1183, Oct. 2018, doi: 10.1080/01691864.2018.1528178.

[12] P. T. Phan, T. T. Hoang, M. T. Thai, H. Low, N. H. Lovell, and T. N. Do, "Twisting and Braiding Fluid-Driven Soft Artificial Muscle Fibers for Robotic Applications," *Soft Robot.*, vol. 9, no. 4, pp. 820–836, Aug. 2022, doi: 10.1089/soro.2021.0040.

[13] R. J. Webster and B. A. Jones, "Design and kinematic modeling of constant curvature continuum robots: A review," *Int. J. Rob. Res.*, vol. 29, no. 13, pp. 1661–1683, Jun. 2010, doi: 10.1177/0278364910368147.

[14] B. A. Jones and I. D. Walker, "Kinematics for multisection continuum robots," *IEEE Trans. Robot.*, vol. 22, no. 1, pp. 43–55, Feb. 2006, doi: 10.1109/TRO.2005.861458.

[15] T. Liu *et al.*, "Iterative Jacobian-Based Inverse Kinematics and Open-Loop Control of an MRI-Guided Magnetically Actuated Steerable Catheter System," *IEEE/ASME Trans. Mechatronics*, vol. 22, no. 4, pp. 1765–1776, Aug. 2017, doi: 10.1109/TMECH.2017.2704526.

[16] A. A. Alqumsan, S. Khoo, and M. Norton, "Robust control of continuum robots using Cosserat rod theory," *Mech. Mach. Theory*, vol. 131, pp. 48–61, Jan. 2019, doi: 10.1016/j.mechmachtheory.2018.09.011.

[17] L. Tang, J. Huang, L. M. Zhu, X. Zhu, and G. Gu, "Path Tracking of a Cable-Driven Snake Robot with a Two-Level Motion Planning Method," *IEEE/ASME Trans. Mechatronics*, vol. 24, no. 3, pp. 935–946, Jun. 2019, doi: 10.1109/TMECH.2019.2909758.

[18] J. Lai, K. Huang, and H. K. Chu, "A learning-based inverse kinematics solver for a multi-segment continuum robot in robot-independent mapping," *IEEE Int. Conf. Robot. Biomimetics, ROBIO 2019*, no. December, pp. 576–582, Dec. 2019, doi: 10.1109/ROBIO49542.2019.8961669.

[19] V. Parenti-Castelli and W. Schiehlen, *ROMANSY 21 - Robot Design, Dynamics and Control*, vol. 47–54. 2016. doi: 10.1007/978-3-319-33714-2.

[20] M. A. Graule, T. P. Mccarthy, C. B. Teeple, J. Werfel, and R. J. Wood, "SoMoGym: A Toolkit for Developing and Evaluating Controllers and Reinforcement Learning Algorithms for Soft Robots," *IEEE Robot. Autom. Lett.*, vol. 7, no. 2, pp. 4071–4078, Apr. 2022, doi: 10.1109/LRA.2022.3149580.

[21] X. Wang and N. Rojas, "A Data-Efficient Model-Based Learning Framework for the Closed-Loop Control of Continuum Robots," *2022 IEEE 5th Int. Conf. Soft Robot. RoboSoft 2022*, pp. 247–254, Apr. 2022, doi: 10.1109/RoboSoft54090.2022.9762115.

[22] M. Wang, X. Dong, W. Ba, A. Mohammad, D. Axinte, and A. Norton, "Design, modelling and validation of a novel extra slender continuum robot for in-situ inspection and repair in aeroengine," *Robot. Comput.*

*Integr. Manuf.*, vol. 67, p. 102054, Feb. 2021, doi: 10.1016/j.rcim.2020.102054.

[23] C. Wang, C. G. Frazelle, J. R. Wagner, and I. D. Walker, “Dynamic Control of Multisection Three-Dimensional Continuum Manipulators Based on Virtual Discrete-Jointed Robot Models,” *IEEE/ASME Trans. Mechatronics*, vol. 26, no. 2, pp. 777–788, Apr. 2021, doi: 10.1109/TMECH.2020.2999847.

[24] Z. Wang *et al.*, “Hybrid Adaptive Control Strategy for Continuum Surgical Robot under External Load,” *IEEE Robot. Autom. Lett.*, vol. 6, no. 2, pp. 1407–1414, Apr. 2021, doi: 10.1109/LRA.2021.3057558.

[25] A. Centurelli, L. Arleo, A. Rizzo, S. Tolu, C. Laschi, and E. Falotico, “Closed-Loop Dynamic Control of a Soft Manipulator Using Deep Reinforcement Learning,” *IEEE Robot. Autom. Lett.*, vol. 7, no. 2, pp. 4741–4748, Apr. 2022, doi: 10.1109/LRA.2022.3146903.

[26] P. Merriaux, Y. Dupuis, R. Boutteau, P. Vasseur, and X. Savatier, “A study of Vicon System positioning performance,” Sensors, vol. 17, no. 7, p. 1591, Jul. 2017, doi: 10.3390/s17071591.